\documentclass[aps,shopacs,twocolumn,pra]{revtex4}
\usepackage{epsfig}
\begin{document}
\title{Screening of qubit from zero-temperature reservoir}
\author{Radim Filip}
\affiliation{
Department of Optics, Research Center for Optics, Palack\' y University,\\
17. listopadu 50,  772~07 Olomouc, \\ Czech Republic}
\date{\today}
\begin{abstract}
We suggest an application of dynamical Zeno effect to isolate a qubit 
in the quantum memory unit against decoherence caused by coupling with the
reservoir having zero temperature. 
The method is based on using an auxiliary casing system that  
mediate the qubit-reservoir interaction and is
simultaneously frequently erased to ground state.
This screening procedure can be implemented in the
cavity QED experiments to store the atomic and photonic qubit states.
\end{abstract}
\maketitle

\section{Introduction}

For efficient quantum information processing, the robust quantum
memories that store information encoded in the state superposition are needed.
The main obstacle in their realization arises from the difficulty to 
isolate a quantum mechanical system from the environment that causes
decoherence effect.
To avoid the decoherence, several strategies have been developed:
the quantum error correction schemes \cite{qec},
feedback implementations \cite{feedback}, the schemes using the
decoherence-free subspace
\cite{dfs,dfsexp}, dynamical decoupling techniques
\cite{Viola98} and the engineering of pointer state methods \cite{eng}.
In this work, we suggest a mechanism to physically isolate the qubit
stored in the memory unit. The basic idea has been inspired by a screening effect
in the electromagnetic field theory, where a device can be screened from
the disturbing effects of the external electromagnetic fields, if it is
closed in the earth-connected metal casing.
To implement this idea in the quantum memories, we use a quantum casing
system that is frequently erased to the ground state, as is
depicted in Fig.~1. As the number of erasing events increases during the time evolution,
the system is better isolated from an action of the zero-temperature reservoir.
From different point of view, the effect 
reminiscents {\em dynamical} Zeno effect \cite{Luis96}, where the system
dynamics can mimic the standard Zeno effect \cite{zeno}.
The possible experimental applications could be found, for example, in the cavity QED
experiments to screen the atomic and photonic qubit, however, it
may be also consider more generally to isolate the system from
the reservoir influence.


To illustrate the properties of the screening method,
we analyze the standard example of decoherence effect:
atomic qubit coupled to the external reservoir on zero temperature, 
for review \cite{Scully97}. 
A two-level atom placed in the free-space,
initially prepared in the pure state superposition
$|\Psi\rangle=\sqrt{1-p}|g\rangle+\exp(-i\psi)
\sqrt{p}|e\rangle$ of the excited state $|e\rangle$ and
ground state $|g\rangle$ and resonantly interacting with a
effectively infinite
reservoir of free-space vacuum modes of electromagnetic
field exhibits {\em exponential} decay of the atomic state.
It can be described by the following master equation in Markovian approximation
\begin{equation}
\frac{d}{dt}\rho=\frac{\gamma}{2}\left(2\sigma\rho\sigma^{\dag}-\sigma^{\dag}
\sigma\rho-\rho \sigma^{\dag}\sigma\right),
\end{equation} 
where $\sigma=|g\rangle\langle e|$ and
$\sigma^{\dag}=|e\rangle\langle g|$
are the deexcitation and excitation
operators of the atomic system. After the reservoir interaction,
the initial state $|\Psi\rangle$ transfers to the reduced density matrix
\begin{eqnarray}\label{marg}
\rho&=&\left(1-P(t)\right)|g\rangle\langle g|+P(t)
|e\rangle\langle e|+\nonumber\\
& &V(t)\left(|g\rangle\langle e|\exp(i\psi)+
|e\rangle\langle g|\exp(-i\psi)\right),
\end{eqnarray}
where $P(t)=P(0)\exp(-\gamma t)$, $P(0)=p$
is probability of the excited state and
$V(t)=V(0)\exp(-\gamma t/2)$, $V(0)=\sqrt{p(1-p)}$ is atomic
interference term.
Thus, the parameters $P(t)$ and $V(t)$ describe
the energy and interference
evolution in the two-level atom.
The fidelity $F=\langle\Psi|\rho|\Psi\rangle$ 
with initial pure atomic state $|\Psi\rangle$ given by
\begin{equation}\label{fidelity}
F(t)=\left(1-P(0)\right)\left(1-P(t)\right)+P(0)P(t)+2V(0)V(t)
\end{equation} 
{\em exponentially} vanishes in course of time,
according to time-behavior of
the depletion coefficients $P(t)$ and $V(t)$.
This fidelity vanishing is one from the main obstacles that inhibit the
efficiency of state preserving in the qubit memories.

\section{Screening procedures}

The idea of screening is based on inserting an auxiliary
casing system, as is depicted
in Fig.~1 and frequent erasing of its state. 
According to the example discussed in the previous Section, 
we have placed the atom with Rabi frequency $\Omega$ in the resonant one-mode high-Q cavity
with damping constant
$\gamma$ \cite{Sachdev84}, as is depicted in Fig.~2.
The casing system is the mode of
electromagnetic field inside the cavity, initially prepared in the vacuum state.
We assume that the off-resonant dissipations described by the damping
constant $\Gamma$ can be neglected in comparison with Rabi frequency $\Omega$ 
and focus on the elimination of decoherence for the resonant
coupling. According to screening idea, we in addition divide the evolution to the $N$ identical
time interval $T=t/N$ by the cavity erasing. 
In quantum theory, an erasure of a single unknown state $|\psi\rangle$
may be thought of as ideal quantum transmission
$|\psi\rangle|0\rangle\rightarrow |0\rangle|\psi\rangle$
with some blank ground state $|0\rangle$ followed by the damping of $|\psi\rangle$ 
to the environment. 
The erasing mechanism can be implemented by an additional
stream of the two-level atoms also resonantly interact with the cavity field, as is depicted in figure, 
and initially prepared in the ground state. The interaction is
adjusted in such a way, that the photon (emitted by the atom)
is quickly extracted in deterministic way
from the cavity. Consequently, the cavity field decouples from the
atomic system and restores the vacuum state in the cavity.
To achieve it, the Rabi frequency $\Omega_{e}$
of the atoms consisting the erasing mechanism must be adjusted $\Omega_{e}\gg
\Omega$. Experimentally, the erasing mechanism can be implemented
analogically to micromaser operation \cite{micromaser},
where the condition $\Omega_{e}\gg \Omega$ can be achieved situating the atom
$A$ near the intensity minimum of standing cavity mode, where the coupling constant
is much smaller than near the maximum. On the other hand, a stream of
atoms consisting the erasing mechanism is focused into
the maximum of standing cavity mode.

Thus, the evolution of screened two-level atom
consists $N$ same sequential parts separated by the erasing events,
in which the atom repeatedly interacts with the damped cavity in the vacuum
state. The evolution between erasing events
can be described in Markovian approximation by
the master equation for atom-field state
\begin{equation}
\frac{d}{dt}\rho=-\frac{i}{\hbar}[H_{I},\rho]+
\frac{\gamma}{2}\left(2a\rho a^{\dag}-a^{\dag}
a\rho-\rho a^{\dag}a\right),
\end{equation} 
where $H_{I}=i\hbar\Omega(a^{\dag}\sigma-\sigma^{\dag}a)$ is interaction
Hamiltonian of the resonant atom-field coupling. Assuming the atom
initially in the state $|\Psi\rangle$ and the cavity mode in the vacuum
state, the master equation can be analytically solved and two
regions of evolution are obtained: for the over-damped case $4\Omega/\gamma<1$, the solution
exhibits the hyperbolic character, which changes for the under-damped case $4\Omega/\gamma>1$ 
in the oscillatory behavior. After tracing
over the cavity field and $N$-times repeated evolution with erasing,
we obtain the reduced density matrix (\ref{marg}) of the two-level atom
having the following coefficients for $\Omega\not= \gamma/4$
\begin{eqnarray}
P_{N}(t)&=&P(0)\exp\left(-\frac{\gamma t}{2}\right)\left[
\frac{\gamma}{4\Lambda}\sinh 2\Lambda\frac{t}{N}+\right.\nonumber\\
& &\left.\frac{\left(
\frac{\gamma}{4}\right)^{2}-\frac{\Omega^{2}}{2}}{\Lambda^{2}}\cosh
2\Lambda\frac{t}{N}-\frac{\Omega^{2}}{2\Lambda^{2}}\right]^{N},\nonumber\\
V_{N}(t)&=&V(0)\exp\left(-\frac{\gamma t}{4}\right)\left[\cosh\Lambda\frac{t}{N}+
\frac{\gamma}{4\Lambda}\sinh\Lambda\frac{t}{N}\right]^{N},\nonumber\\
\end{eqnarray}
where $\Lambda=\sqrt{\left(\gamma/4\right)^{2}-\Omega^{2}}$ and
for $\Omega= \gamma/4$,
\begin{eqnarray}
P_{N}(t)&=&P(0)\exp\left(-\frac{\gamma
t}{2}\right)\left[1+\frac{\gamma}{2N}t+\frac{\gamma^{2}}{16N^{2}}
t^{2}\right]^{N},\nonumber\\
V_{N}(t)&=&V(0)\exp\left(-\frac{\gamma
t}{4}\right)\left[1+\frac{\gamma}{4N}t\right]^{N}.
\end{eqnarray}
Now we arrive to the main result of the paper:
as can be proved,
for every value of the constants $\Omega$ and $\gamma$,
the energy depletion term $P_{N}(t)$ and
interference term $V_{N}(t)$ tend to initial
values $P(0)$ and $V(0)$ as the number $N$ of erasing events increases and consequently,
the fidelity (\ref{fidelity}) 
\begin{equation}\label{fidelity}
F_{N}(t)=\left(1-P(0)\right)\left(1-P_{N}(t)\right)+P(0)P_{N}(t)+2V(0)V_{N}(t)
\end{equation}
approaches unity in the limit
\begin{equation}
\lim_{N\rightarrow\infty}F_{N}(t)=1,
\end{equation}
as is depicted in Fig.~3. Thus the qubit is isolated from the external influence of
zero-temperature.
This effect can be employed as {\em screening} of quantum system from
the reservoir decoherence. The nature of this behavior is
dynamical Zeno effect \cite{Luis96}, which is achieved by to {\em non-exponential}
character of the decoherence induced by the interaction of the atom with
the cavity mode. As the coupling constant $\Omega$ increases, the
non-exponential offset, depicted in Fig.~3, is shorter and we need
implement the erasing mechanism more quickly.
However, number $N$ of erasing
events during the same time interval, depicted in Fig.~4,
increases only sub-exponentially.
Note that the decoherence is the most progressive for unbalanced
superpositions $|\Psi\rangle$ with $P\rightarrow 1$; for this reason, an
unbalanced superposition has been analyzed in Fig.~3 and Fig.~4.


Now we consider different application of the screening for the qubit
state of field mode $|\Psi\rangle=a|0\rangle+b|1\rangle$, where
$|0\rangle$ and $|1\rangle$ are the vacuum and single photon states,
confined in the high-Q cavity. Simultaneously, we will
discuss an influence of the screening on the internal qubit evolution.

Instead of utilizing the standard
one-mode cavity to confined the photon state,
we consider three coupled one-mode cavities in the configuration
depicted in Fig.~5.
We suppose that the mode in the sandwiched cavity is under an internal
interaction described by interaction Hamiltonian $H'$.
We show that an influence of the screening procedure on this
internal interaction is reduced simultaneously
as the erasing is more frequent.
Without loss of generality, we can consider that one side of the sandwiched
cavity has perfect reflectivity and thus, we simplify the discussion to
one-side case. This situation can be described in Markovian approximation
by the following Heisenberg-Langevin equations
\begin{equation}
\frac{d}{dt}a=-\kappa b-\frac{i}{\hbar}[a,H'],\hspace{0.3cm}\frac{d}{dt}b=-\gamma b+\kappa
a+F(t),
\end{equation}
where $a$ is annihilation operator of the screened field mode, $b$ is
annihilation operator of the auxiliary cavity, $F(t)$ is operator
Langevin force of the reservoir, $\kappa$ is coupling constant between
the cavities, $\gamma$ is damping constant and $H'$ is interaction
Hamiltonian of an internal screened mode interaction.
Similarly to the previous example, we assume that erasing event spends a
short-time interval, i.e. the condition $\Omega_{e}\gg \kappa$ is
satisfied.
Assuming that $N$ is sufficiently large to satisfy the conditions
$\kappa,\gamma\ll N/t$, where $t$ is total time of interaction, the
behavior of the screened filed mode between two
erasing events can be described by the short-time approximation.
Then, the total evolution is expressed by iterative relation
\begin{equation}\label{short}
a(t_{n+1})\approx \left(1-\frac{i\Delta t}{\hbar}{\cal D}_{H'}-
\frac{(\kappa\Delta t)^{2}}{2}\right)a(t_{n})-\Xi(\Delta t),
\end{equation}
where $\Delta t=t_{n+1}-t_{n}=t/N$ is time interval between two erasing events,
${\cal D}_{H'}=[...,H']$ is Lie super-operator corresponding to
Hamiltonian $H'$ and $\Xi$ is operator of external reservoir influence
\begin{equation}\label{contr}
\Xi(\Delta t)=b(0)\kappa\Delta t(1-\gamma\Delta t/2)+\kappa
\int_{0}^{\Delta t}\int_{0}^{t'}F(t'')dt''dt',
\end{equation}
which exhibits vanishing normal ordered moments 
\begin{equation}\label{prop}
\langle\Xi^{\dag m}(\Delta t)\Xi^{n}(\Delta t)\rangle=0
\end{equation}
for the auxiliary mode in vacuum state and the zero-temperature reservoir.
Assuming finite number $N$ of erasing events, we can describe the evolution
after $N$ iteration of the relation (\ref{short})
\begin{equation}
a(t)\approx \left(1-\frac{it}{\hbar N}{\cal D}_{H'}-
\frac{(\kappa t)^{2}}{2N^{2}}\right)^{N}a(0)
-\sum_{n=1}^{N}c_{n}(\Delta t)\Xi_{n}(\Delta t),
\end{equation}
where $c_{n}(\Delta t)=\left(1-(\kappa t)^{2}/(2N^{2})\right)^{n}$
are the time dependent coefficients arising by iterative
procedure, $\Xi_{i}(\Delta t)$ are independent external influence (\ref{contr})
satisfying ${\cal D}_{H'}[\Xi_{i}(\Delta t)]=0$.
The term $(\kappa t)^{2}/2N^{2}$ vanishing in the
the limit of erasing events $N\rightarrow\infty$  and the 
the annihilation operator approaches
\begin{equation}
a(t)\approx \exp\left(-\frac{it}{\hbar}{\cal D}_{H'}\right)a(0)
-\sum_{n=1}^{\infty}c_{n}(\Delta t)\Xi_{n}(\Delta t).
\end{equation}
The residual influence of the reservoir is eliminated considering 
(\ref{prop}). Thus, as the number of erasing events $N$ increases, 
all the normal ordered moments 
are protected from the decoherence and consequently, the screened state evolve only under the internal evolution. 
In this way, we may manipulate with the state 
confined in the memory unit.

In this paper, we suggest an application of dynamical Zeno effect \cite{Luis96} to
quantum screening of system qubit from the vacuum reservoir influence.
The screening exhibits independently on the coupling and damping parameters and can be applied
in both the over-damped and under-damped cases. 
In addition, the internal system evolution is simultaneously
preserved and we thus can operate with quantum state saved in the memory.
Two possible experimental
implementations have been pointed up, for the two-level atom and for photonic
qubit state stored in the cavities.

\medskip
\noindent {\bf Acknowledgments}
I would like to thank Jarom\' \i r Fiur\' a\v sek and L. Mi\v sta Jr. for the fruitful discussions.
The work was supported by the project LN00A015
of the Ministry of Education of Czech Republic.


\begin{figure}
\vspace{0cm}
\vspace{0cm}
\caption{Screening of the qubit state: A -- direct qubit interaction with
reservoir with damping constant $\gamma$, B -- 
screening of reservoir interaction by frequently erased casing qubit coupled to the system.}
\end{figure}

\begin{figure}
\vspace{0cm}
\vspace{0cm}
\caption{Schematic setup of the atomic state screening: A -- screened
atom, $\gamma$ -- cavity damping constant, $\Gamma$ -- atomic damping
constant, $\Omega$ -- Rabi frequency of atom-cavity coupling,
$\Omega_{e}$ -- Rabi frequency of erasing interaction.}
\end{figure}
\begin{figure}
\vspace{0cm}
\vspace{0cm}
\caption{The slowing down of fidelity degradation by the screening
effect: the atom confined in damped cavity
for over-damped (under-damped) behavior.
For illustration, we assume the dissipative constant $\gamma$
of the cavity same as for the direct atom-reservoir interaction case.}
\end{figure}
\begin{figure}
\vspace{1cm}
\vspace{0cm}
\caption{Number of the erasing events which is needed to
preserve the qubit state ($p=0.9$) for
$\gamma t=\pi$.}
\end{figure}
\begin{figure}
\vspace{0cm}
\vspace{0cm}
\caption{Schematic setup of the field state screening:
$\gamma$ -- cavity damping constant, $\kappa$ -- cavity coupling
constant, $\Omega_{e}$ -- Rabi frequency of erasing interaction.}
\end{figure}

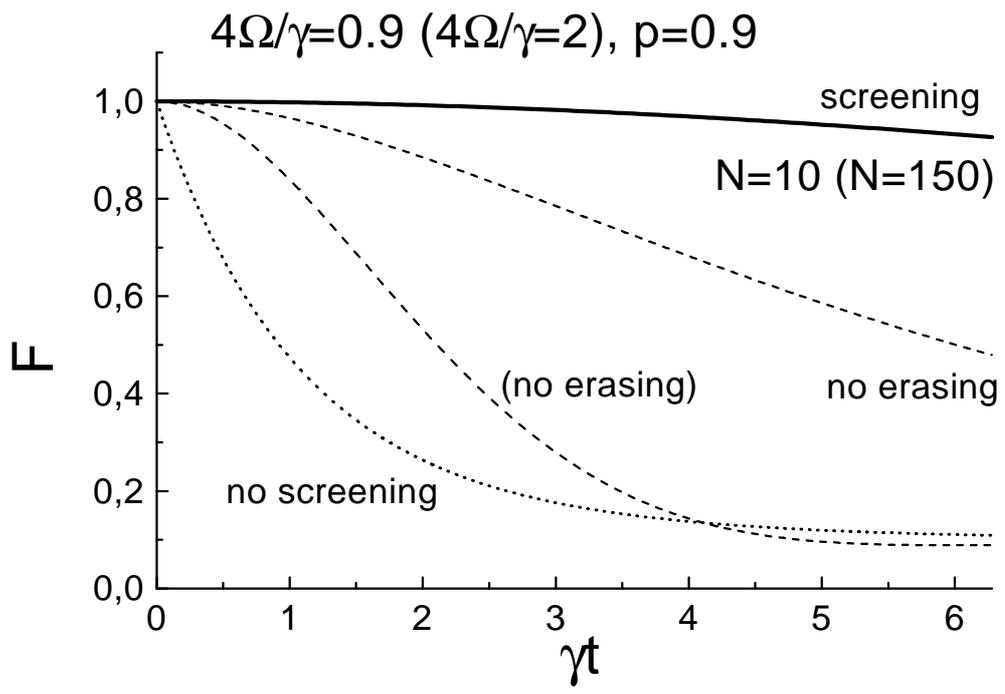
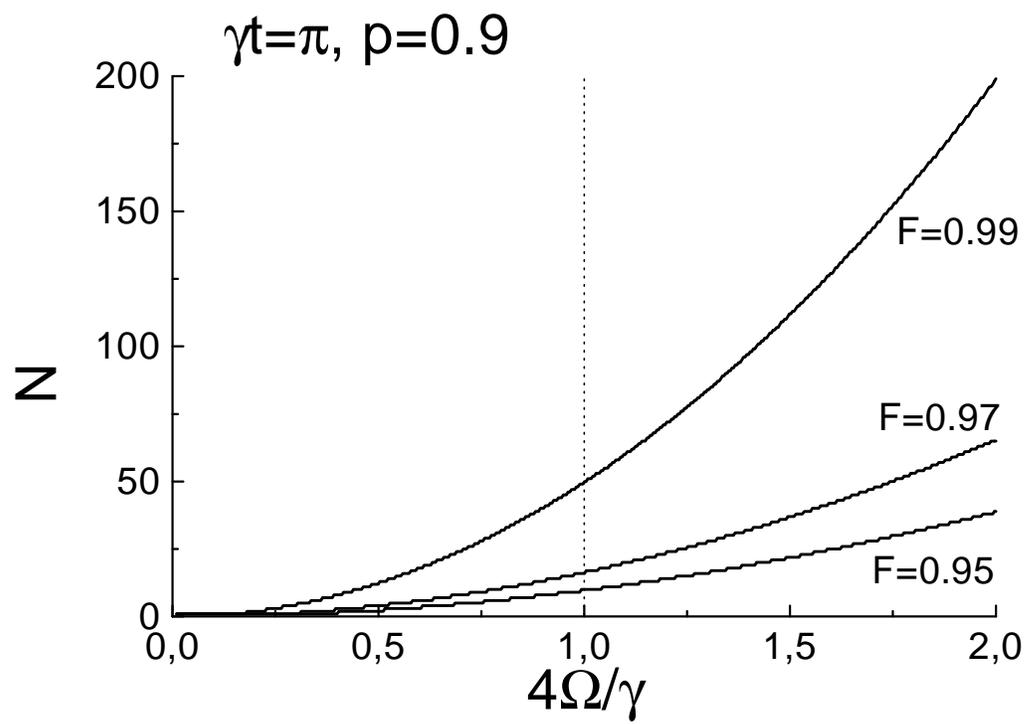
\end{document}